\documentclass[prl,twocolumn,showpacs,floatfix,amsmath,amssymb,superscriptaddress] {revtex4}
\pdfoutput=1
\usepackage{graphicx}
\usepackage{times}
\usepackage{amsmath}
\usepackage{dcolumn}
\newcommand{\beq}{\begin{equation}}
\newcommand{\eeq}{\end{equation}}
\newcommand{\bea}{\begin{eqnarray}}
\newcommand{\eea}{\end{eqnarray}}
\newcommand{\be}{\begin{equation}}
\newcommand{\ee}{\end{equation}}

\def\CsCuCl{$\text{Cs}_2\text{CuCl}_4$}
\def\CsCuBr{$\text{Cs}_2\text{CuBr}_4$}

\def\DM{\text{Dzyaloshinskii-Moriya}}
\def\beq{\begin{equation}}
\def\eeq{\end{equation}}
\def\bea{\begin{eqnarray}}
\def\eea{\end{eqnarray}}

\begin{document}

\title{Spin-current order in anisotropic triangular antiferromagnets}

\author{Andrey V. Chubukov}
\affiliation{Department of Physics, University of Wisconsin, Madison,
WI 53706}
\author{Oleg A. Starykh}
\affiliation{Department of Physics and Astronomy, University of Utah, Salt Lake
City, UT 84112}

\date{\today}

\begin{abstract}

We analyze instabilities of the collinear
 up-up-down  state of
 a two-dimensional quantum spin-$S$
spatially anisotropic  triangular lattice antiferromagnet in a magnetic field.
 We find, within large-$S$ approximation, that near the end point of the plateau, the collinear state becomes unstable due to
 condensation of
  two-magnon bound pairs rather than single magnons.
  The two-magnon instability leads to
   a novel  2D vector chiral phase
 with alternating spin currents
 but no
  magnetic order in the direction transverse to the field.
This phase
breaks a discrete $Z_2$ symmetry but preserves a continuous $U(1)$
one of rotations about the field axis.
It possesses orbital antiferromagnetism and displays a magnetoelectric effect.
\end{abstract}

\maketitle

%%%%%%%%%%%%%%%%%%%%%%%%%%%%%%%%%%%%%%%%%%%%%%%%%%%%%%%%%%%%%%%%%%%%%%

{\underline{\emph{Introduction}.}~~~~ The field of frustrated quantum magnetism has witnessed a remarkable revival of interest in recent years due to
rapid progress in the fabrication and characterization of new materials and a multitude of theoretical ideas about competing orders and
new quantum states of matter \cite{leon}. Studies of two-dimensional (2D) quantum triangular lattice
antiferromagnets with spatially anisotropic exchange, such as \CsCuCl~and \CsCuBr,
are of particular interest because of their surprisingly rich
 phase diagrams
  in a magnetic field~\cite{tokiwa06,takano08}
   which includes novel quantum states  which have no classical analogs and
   display a wealth of properties which are highly sought after for  applications.
   The  large number of different phases involved, which reaches 9 in the case of \CsCuBr \cite{takano08},
   reveals a highly complex   interplay between quantum fluctuations and 
  anisotropy of the interactions.
 
 One of the best understood  phases
  of a frustrated spin system in a magnetic field
is a collinear state
 with a fixed,  field-independent  magnetization equal to
 exactly
 1/3 of the saturation value.
In this  state, known as the up-up-down (UUD),  two spins in each triangle point up and one points down.
This quantum state preserves continuous $U(1)$ symmetry
of rotations about the field direction and has finite gaps in all spin excitations \cite{chubukov91}.
The UUD state is  similar to plateau states in quantum Hall effect, although, unlike them, it
spontaneously breaks lattice translational symmetry. An extension of the UUD state with unbroken 
  translational symmetry has been proposed  
   theoretically but not yet found experimentally \cite{misguich2001,alicea2007}.

\begin{figure}
\begin{center}
  \scalebox{0.99}{\includegraphics[width=\columnwidth]{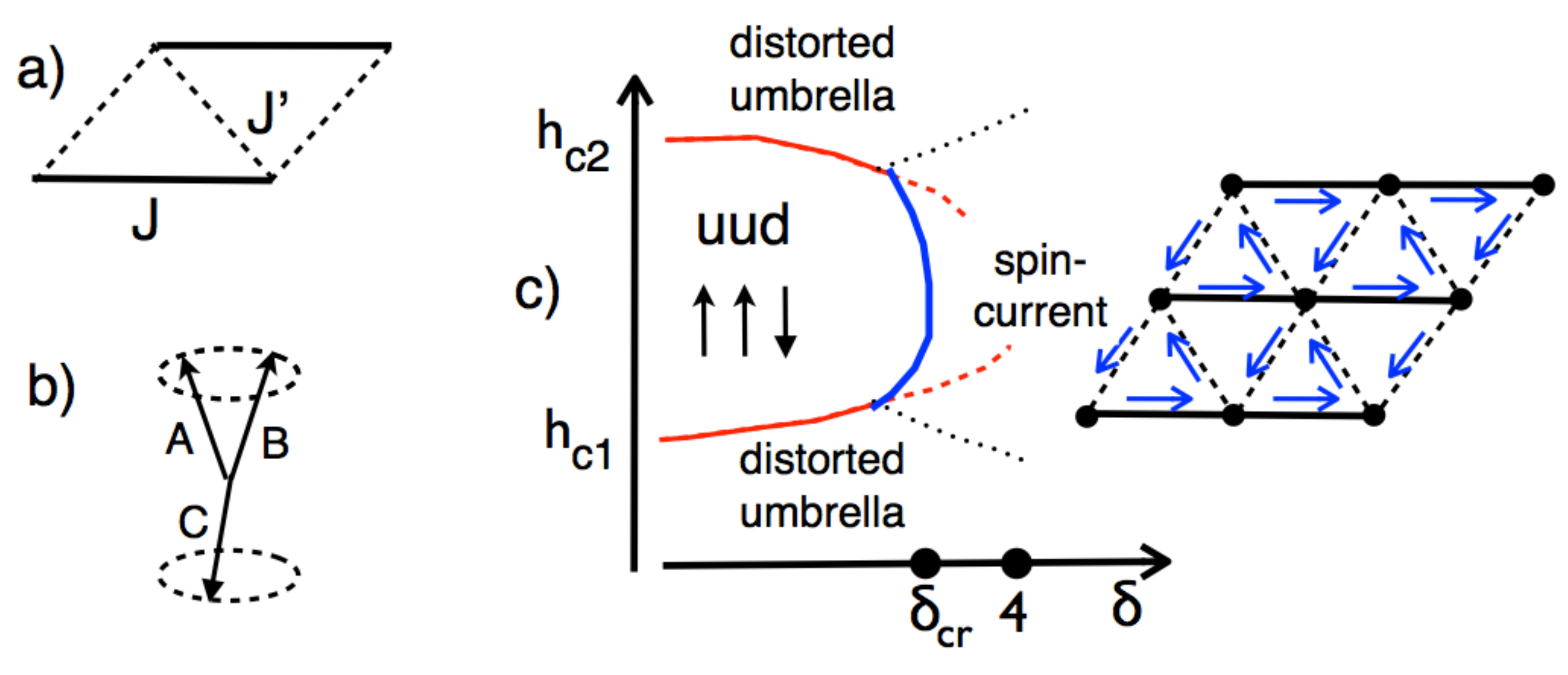}}
  \end{center}
\caption{(Color online) (a) Anisotropic triangular lattice with exchanges $J$ and $J'$. (b)
 Distorted umbrella state. (c) Schematic phase diagram of the
model in the vicinity of the UUD end-point at $\delta=4$. Thin solid (red) lines mark
 single-particle instabilities of the UUD state at $h_{c1, c2}(\delta)$.
Thick solid (blue) line is the two-particle instability line
  towards a spin-current state, which emerges at  $\delta > \delta_{\rm cr}$, and dotted
   (black)  lines indicate phase transitions between the umbrella and the spin-current state.
    Dashed (red) line indicates a would-be single-particle instability, which is pre-empted by the
 two-particle instability.
(Blue) arrows in the insert on the right show the arrangement of spin currents.}
\label{fig:phase}
\end{figure}

In a classical isotropic 2D Heisenberg systems with nearest exchange $J$, the UUD phase is the ground state for just one value of the external field $h = 3J$ (
1/3 of the saturation field $h_{\rm sat} = 9J$).
At all other fields spins order in a non-collinear fashion.
 In an anisotropic lattice with exchanges $J$ and $J'$ (see Fig. 1), a
 non-collinear order wins for all fields, so that classically UUD phase is never a ground state.
  For quantum systems, the situation is different as quantum fluctuations favor a collinear spin structure
   and compete with classical fluctuations~\cite{chubukov91,alicea,alicea2}.
   In the isotropic case, quantum
   fluctuations stabilize
  the  UUD phase with gapped spin-wave excitations in a finite interval of $h$
   with the width of order $1/S$.
  In an anisotropic case, the
  width of the UUD phase is determined by
  the competition between $1/S$, which measures the strength of quantum
   fluctuations, and the degree of antisotropy of exchange interactions
   $(1-J'/J)$ (Ref.~\cite{alicea}). The dimensionless parameter, which determines
   the UUD width
   relative  to its value in the isotropic case, is $\delta = (40/3) S (1-J'/J)^2$ (we use the same numerical factor  as in~\cite{alicea}).
    The UUD phase persists up to a finite anisotropy $\delta_{\rm cr} = 4$, see Fig.~\ref{fig:phase}.
       The boundaries of the UUD phase have been determined from the local stability analysis ~\cite{alicea}
   as the values of $h$ at which spin-wave dispersion softens.
   Of the two low-energy spin-wave branches, one softens at the lower boundary of the UUD phase
       and another at the upper boundary.
      Near the critical $J'/J$, both spin-wave instabilities occur at finite momenta, and each leads to a chiral, non-coplanar state
      (often called a
       distorted umbrella),
      in which $\langle{\bf S}_{\bf r}\rangle$ has finite components along both directions perpendicular to the field~\cite{alicea,zhit}
      (see Fig.~\ref{fig:phase}).

The analysis of the same model for $S=1/2$,  however, found very different
  states surrounding the UUD plateau near its end point, which for $S=1/2$ extends all way to $J'=0$ \cite{sdw2012}.
  These states are collinear spin-density wave (SDW) states, with
      incommensurate spin modulations along the field direction but {\em no} long-range order in the transverse direction \cite{sdw2012}.
    This
    discrepancy
    poses the question whether the phase diagram
   for $S=1/2$ is qualitatively different from the one at large $S$, or
   the ground states surrounding the UUD phase are
   different from
   the ones predicted by spin-wave theory even for large $S$.

   In this work we re-visit the large $S$ analysis of the UUD
    state
    and show that the spin-wave phase diagram is
  incomplete {\it for any $S$}.
  We show that,
   prior to a single-magnon instability,
 the system undergoes a
  pairing instability,
in which the two-particle collective mode, made of magnons from the two low-energy branches,
 softens at zero total momentum of the pair. As a result,
 the actual instability
  near the end point of UUD phase
 is  towards the uni-axial state with no magnetic order in the transverse direction,
 similar to the situation for $S=1/2$.
  We solve the ``gap" equation for the two-magnon order parameter
   and show that it is purely {\it imaginary}.
   Such order parameter breaks a discrete $Z_2$ symmetry and gives rise to a bond-nematic state
  with non-zero vector and scalar chiralities within a single triangle of spins:
 $\langle{\bf S}_A \cdot {\bf S}_B \times {\bf S}_C\rangle \neq 0$ and $\langle {\bf S}_A \times {\bf S}_B\rangle = \langle{\bf S}_B \times {\bf S}_C\rangle
  = \langle{\bf S}_C \times {\bf S}_A\rangle \neq 0$  (vector and scalar chiralities are proportional to each other since
   the total magnetization $M=\langle S_z\rangle$ is finite).
 Such a state supports circulating spin currents (Fig.~\ref{fig:SC}) and we label it a {\em spin-current} state (SC).
   We present the modified large-$S$ phase diagram of the model in Fig.~\ref{fig:phase}.

Experimental signatures of a SC state are rather peculiar.  First, it
  exhibits a magneto-electric effect because both spin current and electric field are odd  under spatial reflections and
 couple linearly~\cite{bal_nag}. As a result, spin-wave excitations of the SC state depend linearly on $E$.
 Second, orbiting spin currents generate charge currents, which in turn
produce staggered magnetic moments, which can be measured by NMR and $\mu$SR~\cite{batista_09}.

{\underline{\emph{The model}.}}~~~~ We consider a system of localized spins on
 an anisotropic triangular lattice with Heisenberg nearest-neighbor interactions $J$ and $J'$,
  subject to an external field ${\tilde h} = 2\mu_B H_z$:
\be
{\cal H} = \sum_{\bf r} \left(  J  {\bf S}_{\bf r} {\bf S}_{\bf r+a_x} + J' \sum_{j=1,2}{\bf S}_{\bf r} {\bf S}_{\bf r+a_j} -
\tilde{h}  S^z_{\bf r}\right),
\label{1}
\ee
where ${\bf a}_{1,2} = a (1/2, \pm \sqrt{3}/2)$ connects spins on neighboring chains, and $a$ is the lattice constant.
For convenience, we rescale $\tilde{h} = h S$ and use $h$ for the field.
 The saturation field, above which the magnetization $M$ reaches maximum possible value $M_{\rm sat} = S$,
is given by $h_{\rm sat} = (2J + J')^2/J$.  We are interested in the behavior of the system
 near $h_{\rm sat}/3$, where quantum fluctuations
  win over classical fluctuations and stabilize UUD phase
 in a finite range of fields.
  In the isotropic case, $J'=J$, the UUD phase exists
  in a field range
   between $h_{c1} = (h_{\rm sat}/3) (1 -0.5/2S)$ and $h_{c2} = (h_{\rm sat}/3) (1 +1.3/2S)$.
   In the anisotropic case, $J' < J$, the width of the UUD state decreases and eventually vanishes
    at $\delta_{\rm cr} =4$, which defines
    $J'_{\rm cr} = J(1 - \sqrt{3/10S})$

The excitation spectrum of the UUD phase
 at $\delta \leq 4$
can be straightforwardly obtained by using a three-sublattice representation for two spin-up
and one spin-down
 sublattices and introducing \cite{alicea,suppl} three sets of Holstein-Primakoff bosons,
  $a, b$, and $c$.
 One  of the three spin-wave branches describes
 the precession of the total magnetization, has energy of the order $h_{\rm sat}/3$, and is irrelevant to our analysis.
The other two branches,
denoted $d_{1(2), {\bf k}}$ below,
 describe low-energy excitations.
  Explicitly,
   \begin{equation}
  {\cal H}^{(2)}_{\text{uud}} = S\sum_{\bf k} \Big( \omega_1 d^\dagger_{1,\bf k}d_{1,\bf k}
+ \omega_2 d^\dagger_{2,\bf k} d_{2,\bf k}\Big),
\label{eq:Huud}
\end{equation}
where at small ${\bf k}$
\begin{eqnarray}
\label{eq:uud5}
 \omega_{1,2}({\bf k}) &=&
 \pm \left(h-h_0-\frac{1}{5S}J-\frac{3}{4}J{\bf k}^2\right) + \frac{3J}{20S} Z_{\bf k}, \\
 Z_{\bf k} &=& \sqrt{9 + 10 S (6 {\bf k}^2 -3\delta k_x^2 + 10 S {\bf k}^4)},
\label{eq:uud6}
\end{eqnarray}
 and $h_0 = J + 2 J'$.
The excitation $d_{1,\bf k}$ softens at the lower boundary of the UUD phase, at $h = h_{c1}(\delta) = h_{\rm end} - 9J/(40S) \sqrt{(4-\delta)/3}$, where
$h_{\rm end} = h_0 (1 +17/(120S))$. The softening happens at a finite momenta
 $\pm {\bf k}_1  = (\pm k_1,0)$, where $k_1 \approx (3/(10S))^{1/2}(1+\sqrt{(4-\delta)/12})$. The excitation $d_{2,\bf k}$ softens  at the upper boundary $h = h_{c2}(\delta) =
h_{\rm end} + 27J/(40S) \sqrt{(4-\delta)/3}$,  at momenta $\pm {\bf k}_2  = (\pm k_2,0)$, where $k_2 = (3/(10S))^{1/2}(1-\sqrt{(4-\delta)/12})$.
 The spin-wave softening at either $h_{c1}(\delta)$ or $h_{c2}(\delta)$ signals
 condensation of one-magnon excitations.  A Ginzburg-Landau-type analysis
 shows~\cite{alicea} that condensation
  spontaneously
  breaks
 $Z_2$ symmetry between degenerate minima at $\pm {\bf k}_{1}$ and $\pm {\bf k}_2$.
 As a result, one-magnon condensation gives rise to an
  incommensurate spiral order with spontaneously broken
 $O(2) \times Z_2$ symmetry and a finite non-coplanar long-range order $\langle{S_{\bf r}^{x,y}}\rangle \neq {\bf 0}$.

 At the end-point of the plateau $\delta =4$, $h_{c1} = h_{c2} = h_{\rm end}$,
both spin-wave branches touch zero
simultaneously at $\pm {\bf k}_0 = (\pm k_0,0)$, where
 $k_0= \sqrt{3/(10S)}$. The presence of four soft modes
 leads to a variety of possible
 non-coplanar chiral orders with non-zero
 $\langle{S_{\bf r}^{x,y}}\rangle$.
 However, we show below that instead
 the system undergoes a pre-emptive pairing instability into a state with no
transverse order, $\langle S_{\bf r}^{x,y} \rangle =0$, but
 nonetheless  with a finite chirality $\langle\hat{z}\cdot{\bf S}_{\bf r} \times {\bf S}_{\bf r'}\rangle \neq 0$.

{\underline{\emph{Magnon pairing}.}}~~~~To analyze a
 possibility of a bound state of two magnons, we need to include magnon-magnon interaction.
The derivation of the interaction Hamiltonian is lengthy but straightforward:
  one has to express two-magnon interaction Hamiltonian ${\cal H}_{\rm uud}^{(4)}$,
 originally written in terms of $a_{\bf k}, b_{\bf k}$ and $c_{\bf k}$ bosons, in terms of the low-energy eigen-modes
 $d_{1,{\bf k}}$ and $d_{2,{\bf k}}$ from Eq. (\ref{eq:Huud}).
 The full transformation is given in \cite{suppl}. Near momenta
$\pm {\bf k}_0$, which are mostly relevant to the pairing problem, this transformation simplifies to
\begin{eqnarray}
&&a_{\bf k} = \frac{f({\bf k})}{\sqrt{2}}(e^{i s_{\bf k}} d_{1,{\bf k}} - e^{- i s_{\bf k}} d_{2, -{\bf k}}^\dagger),\nonumber\\
&&b_{\bf k} = -\frac{f({\bf k})}{\sqrt{2}}(e^{-i s_{\bf k}} d_{1,{\bf k}} + e^{ i s_{\bf k}} d_{2, -{\bf k}}^\dagger),\nonumber\\
&&c_{\bf k} = f({\bf k})(d_{2,{\bf k}} - e^{ i 2s_{\bf k}} d_{1, -{\bf k}}^\dagger).
\label{eq:abc}
\end{eqnarray}
where
$f({\bf k}) = \sqrt{k_0} [(k_x \pm k_0)^2 + k_y^2 + (1- \delta/4) k_0^2]^{-1/4}$
and $s_{\bf k} = \pi ~{\rm sign}(k_x)/4$.

Consider first $\delta <4$, when only one boson becomes soft at either $h_{c1}$ or $h_{c2}$,
 while other remains massive and can be neglected.
  For concreteness,
consider the vicinity of  $h_{c1}$,
where $d_1$ excitation softens. The magnon-magnon
 pairing
 interaction involving only $d_1$ bosons is
\be
  {\cal H}^{(4)}_{d_1d_1} =  \frac{8 (J + 2 J')}{(4-\delta)} \frac{3}{N} \sum_{p,q}
  d^\dagger_{1,{\bf k}_1+{\bf p}}d^\dagger_{1,{-\bf k}_1-{\bf p}} d_{1,{\bf k}_1+{\bf q}}d_{1,{-\bf k}_1-{\bf q}}
\label{mon_1}
\ee
 This interaction is obviously strongly repulsive and does not give rise to
  a  bound state. The same holds for $d_2$ mode near $h_{c2}$.
  As a result, one-magnon condensations at $h_{c1}$ and $h_{c2}$
 are the true instabilities, and the system develops a
 non-coplanar
 spiral order at $h \geq h_{c2}$ and $h \leq h_{c1}$.

 For $\delta \approx 4$,
 the situation is different. Magnon-magnon interactions within
 $d_1$ or $d_2$ sectors are still repulsive, but now we also
  have interaction between $d_1$ and $d_2$ bosons,
 both of which are gapless at $\pm {\bf k}_0$.
  The $d_1-d_2$ interaction
   with zero total momentum has
  two relevant terms: one describes "normal" $2 \to 2$ process with simultaneous creation and annihilation of
  $d_1$ and $d_2$ bosons,
   the other describes "anomalous" $4 \to 0$ and $0 \to 4$ processes with simultaneous creation or annihilation of
  two $d_1$ and two $d_2$ bosons.  We find that the strongest pairing
  interaction involves momentum transfer $\pm 2k_0$ for each of the bosons involved.
   The corresponding interaction reads
  \bea
   &&{\cal H}_{d_1d_2}^{(4)} = \frac{3}{N} \sum_{p,q} \Phi (p,q) \Big(
   d^\dagger_{1,{\bf k}_0+{\bf p}}d^\dagger_{2,{-\bf k}_0-{\bf p}} d_{1,{-\bf k}_0+{\bf q}}d_{2,{\bf k}_0-{\bf q}}
   %\}
   \nonumber \\
  &&
  %\{
   - d^\dagger_{1,{\bf k}_0+{\bf p}}d^\dagger_{2,{-\bf k}_0-{\bf p}} d^\dagger_{1,{-\bf k}_0+{\bf q}}d^\dagger_{2,{\bf k}_0-{\bf q}} \Big) + \text{h.c.}
   % \Big)
\label{mon_2}
\eea
where $p$ and $q$ are much smaller than $k_0$, and the vertex
\be
\Phi (p,q) = -(J + 2J') f^2(p) f^2(q) \to - (J + 2J')\frac{k^2_0}{|{\bf p}||{\bf q}|}
 \label{mon_4}
 \ee
where $f(p)$ was introduced after Eq. (\ref{eq:abc}), and the limit stands for $\delta \to 4$.
 The pairing interaction with small momentum transfer,
${\tilde \Phi} (p,q) d^\dagger_{1,{\bf k}_0+{\bf p}}d^\dagger_{2,{-\bf k}_0-{\bf p}} d_{1,{\bf k}_0+{\bf q}}d_{2,{-\bf k}_0-{\bf q}}$,
has a much smaller ${\tilde \Phi} (p,q)$ which remains finite
 in the limit $p, q \to 0$. Such interaction is then irrelevant for our analysis.

Now observe that the sign of $2 \to 2$ term is negative, while the one of $4 \to 0$ term is positive. The negative sign
 of the $2\to 2$ term implies that the ``normal'' interaction between $d_1$ and $d_2$ bosons is attractive and favors a pairing with
 %OS aug7
 \bea
 &&F_{{\bf k}_0} (p) = \langle d_{1,{\bf k}_0+{\bf p}}d_{2,-{\bf k}_0-{\bf p}}\rangle  =  F_{-{\bf k}_0} (p) =\\
&& =\langle d_{1,-{\bf k}_0+{\bf p}}d_{2,{\bf k}_0-{\bf p}}\rangle  = 
 \frac{{\tilde \Upsilon} f^2({\bf p})}{\omega_{1}({\bf k}_0 + {\bf p}) + \omega_{2}({\bf k}_0 + {\bf p})} \to \frac{{\tilde \Upsilon}}{{\bf p}^2}. \nonumber 
 %= F_{-{\bf k}_0} (p).
 \eea
The positive sign of the
 $4 \to 0$ term does not allow the solution with real $\tilde\Upsilon$ (the corresponding coupling constant vanishes), but instead favors a solution with
 imaginary $\tilde\Upsilon = i \Upsilon$.
 For such solution
  the pairing vertex which couples to $4\to 0$ term has opposite sign compared to the vertex which couples to $2\to 2$
 term, and this extra sign change compensates the sign difference between $2\to 2$ and $4 \to 0$ interactions.
Note that since the Hamiltonian \eqref{mon_2} does not conserve
 the number of bosons,  the order parameter does not possess a $U(1)$ phase
symmetry. In practice, this implies that the gap equations for real and imaginary $\Upsilon$'s are different.
 And, in fact, the symmetry that is
spontaneously broken at the transition is $Z_2$, corresponding to the sign of $\Upsilon$.

 For $\tilde\Upsilon = i {\Upsilon}$, the linearized ``gap" equation reads at $\delta=4$,
 \be
 \Upsilon = \frac{6 \Upsilon}{NS} \sum_p \frac{(J + 2J') k^2_0}{{\bf p}^2} \frac{1}{\omega_{1}({\bf k}_0 + {\bf p}) + \omega_{2}({\bf k}_0 + {\bf p})} .
 \label{mon_5}
 \ee
 Substituting the dispersions, we find
 \be
  1 =  \frac{1}{S} \frac{3}{N} \sum_p \frac{k_0}{|{\bf p}|^3}.
  \label{mon_3}
  \ee
  It is important that the integrand scales as $1/|{\bf p}|^3$, so that the 2D integral over ${\bf p}$ diverges
    and overcomes the smallness of $1/S$ in the pre-factor.
   In $1/|{\bf p}|^3$, one power of $1/|{\bf p}|$ comes
  from the
    dispersion  and the other two powers
    are due to the divergence of the coherence factor  $f(p)$ at $p\to 0$.
  Away from $\delta =4$,
   $|{\bf p}|$ is replaced by $(|{\bf p}|^2 + (1-\delta/4) k^2_0)^{1/2}$, and the integral in the r.h.s of (\ref{mon_3}) behaves as
   $1/\sqrt{4-\delta}$.
  Collecting powers of $1/S$, we find that
   a nonzero $\Upsilon$ emerges at $\delta_{\rm cr} = 4 - O\left(1/S^2\right)$.

For completeness, we also analyzed possible pairing with the total momentum $\pm 2 {\bf k}_0$, but found that
there is no enhancement of the kernel of the gap equation by coherence factors and, hence, no instability at large $S$.

{\underline{{\em Spin-current order.}}~~~~The two-magnon instability does not lead to a conventional spin order in the direction perpendicular to the field because
 $\langle d_{1,k}\rangle =\langle d_{2,k}\rangle=0$.
 $F_{{\bf k}_0} (p)\sim \Upsilon$
 does not lead to modulations of $S^z_{\bf r}$ or the bond order because the condensate does not contribute to
 magnon density or to $\langle {\bf S}_A \cdot {\bf S}_B\rangle$~\cite{suppl}.
  However, one can easily verify that for each triangle we now have
 $\langle \hat{z}\cdot {\bf S}_A \times {\bf S}_C\rangle = \langle \hat{z}\cdot {\bf S}_C \times {\bf S}_B\rangle  =
  \langle \hat{z}\cdot{\bf S}_B \times {\bf S}_A\rangle \propto \Upsilon$,
  which implies a finite vector chirality and orbital spin currents which run in opposite directions in neighboring triangles, Figure~\ref{fig:SC}.
  Note that the sign of Ising order parameter $\Upsilon$ determines the sense of spin current circulation.
  In our case vector chirality generates a non-zero scalar chirality  $\langle {\bf S}_A \cdot {\bf S}_B \times {\bf S}_C\rangle \sim \Upsilon$ as well,
  because of the finite magnetization $M$ along the $z$ (magnetic field) axis.
  For triangles separated by distance ${\bf r}$,
  ${\hat z} \cdot \langle{\bf S} (0) \times {\bf S} ({\bf r})\rangle$ scales as 
  %OS aug2
  $\Upsilon \cos({{\bf k}_0 {\bf r}}) ~e^{ - r k_0 \sqrt{1-\delta/4}}$ ~\cite{suppl}.

 \begin{figure}
\begin{center}
  \scalebox{0.8}{\includegraphics[width=\columnwidth]{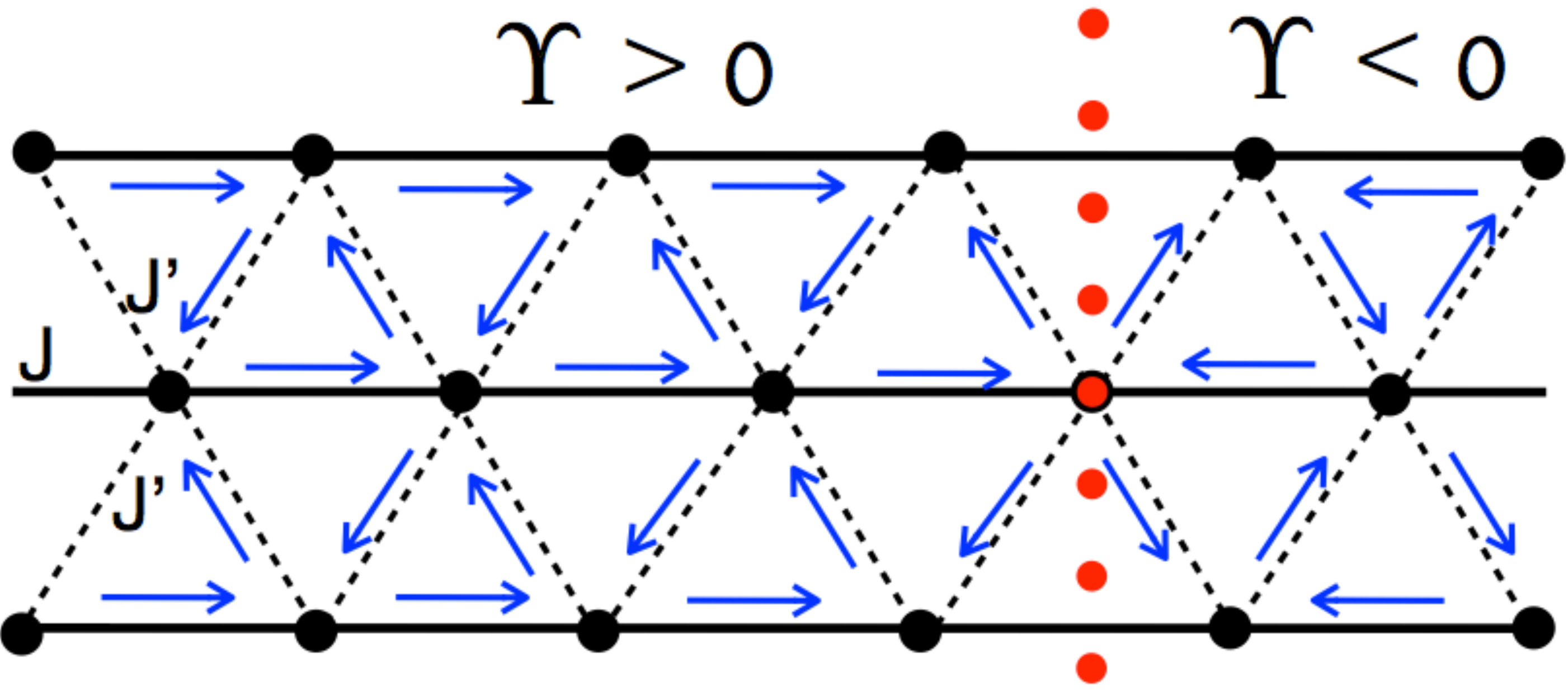}}
  \end{center}
\caption{(Color online) The structure of spin currents in the SC state.
 The  domain wall, denoted by vertical (red) dotted line, separates
   domains with opposite chirality $\Upsilon$.
   }
\label{fig:SC}
\end{figure}

  A  spin-current (SC)
  order in dimensions $D >1$ is normally associated with non-coplanar spin ordering when the spins
  spontaneously select the direction of rotation in the XY plane.
  Remarkably, in our case the SC order appears in the absence of the standard spin order in the XY plane.

 The emergence of the SC order can be thought of as spontaneous generation of Dzyaloshinskii-Moria (DM) interaction. Indeed, the
 interaction Hamiltonian \eqref{mon_2} can be written as
  ${\cal H}_{d_1d_2}^{(4)} = -(9J/N) {\cal H}^{\rm DM}_{k_0} ~{\cal H}^{\rm DM}_{-k_0}$, where \cite{suppl}
 \bea
&& {\cal H}^{\rm DM}_{\pm k_0} = \frac{1}{6S} \sum_r \hat{z}\cdot {\bf S}_{\bf r} \times \left({\bf S}_{\bf r+ a_1} + {\bf S}_{\bf r+ a_2}\right) \nonumber \\
&&=  i \sum_{{\bf k}\in \pm {\bf k}_0} f^2_k \left(d_{1,{\bf k}} d_{2,-{\bf k}} - d^\dagger_{1,{\bf k}} d^\dagger_{2,-{\bf k}}\right) .
\label{wed_2}
\eea
As a result,
 the development of a non-zero $\Upsilon$ can be viewed as the appearance of
 Dzyaloshinskii-Moria interaction
 $ D ({\cal H}^{\rm DM}_{k_0} + {\cal H}^{\rm DM}_{-k_0})$, with
 $D \sim \Upsilon$.
This observation
 helps to understand magneto-electric effect in the SC state: because
 $D$ is a pseudoscalar, it couples linearly
to an electric field $E$, i.e., $D = D_0 + D_1 E + ...$. As a result,
spin-wave excitations of the SC phase depend linearly on $E$.

SC order has been previously explored in 1D spin ladders \cite{shura,kolezhuk,akira-J1J2} and
 was suggested for a frustrated Heisenberg model in 2D~\cite{chandra,lauchli}.
There, however, a SC state is a spiral state, in which a continuous $U(1)$ symmetry is restored by strong quantum fluctuations \cite{lauchli}.
 In our case spiral states are present in the phase diagram away from the end-point of the UUD phase, while
 the SC state
 emerges as a result of a pre-emptive two-magnon instability
 rather than due to divergent one-magnon fluctuations.
 Our two-magnon instability (which necessary leads to an imaginary order parameter) is also
  fundamentally
  different from two-magnon instabilities with real order parameter which lead to a spin-nematic order,
either on a site or on a bond \cite{andreev,chubukov,akira,sudan,mzh,syromyatnikov}. Such
  order generally occurs in systems with ferromagnetic exchanges at least on some of the bonds,
 when there is an
  attractive interaction between magnons.
Here, all exchange couplings are antiferromagnetic, and magnon-magnon interaction is repulsive.
Our pairing of magnons from different branches is conceptually
 similar to the inter-pocket pairing in multi-band fermionic systems, such as Fe-based superconductors with only electron pockets~\cite{khodas}.

The phase diagram near the end point of UUD state has been recently analyzed in ~\cite{zhit} in a
self-consistent semiclassical formalism. This method, however, does not allow for the analysis of two-particle instabilities.

{\underline{{\em {Comparison with SDW state}.}}~~~~
  Although our analysis uses $1/S$ expansion, it is nevertheless instructive to
 compare symmetry properties of our spin-current state with that of a collinear SDW state observed for $S=1/2$ near the end point of the UUD phase.
 Like we said, spin-current state is much closer to SDW state than a spiral state (the result of one-magnon condensation) because both spin-current and SDW states
 preserve $U(1)$ symmetry of rotations about the field direction. But the two states do differ as SDW state has no chiral order \cite{sdw2012}.
It may be that $S=1/2$ is simply special and non-chiral SDW state is only present at $S=1/2$.
But it also may be that the two-magnon instability, which we found, is only  a `tip of the iceberg', and the two-magnon condensation triggers the
 development of multi-magnon condensates at some $\delta >\delta_{\rm cr}$, which in turn changes the properties
 of the spin-current state. This last possibility is inspired by the observation that SDW state is incommensurate and that the
 UUD-SDW transition for $S=1/2$ is a commensurate-incommensurate transition \cite{sdw2012}. Such transition
occurs via a proliferation of solitons -- strings of displaced spins
which are shifted from their equilibrium UUD pattern.  Since changing the direction of a single spin $S$ requires $2S$ magnons, a
proliferation of solitons implies condensation of $2S$ magnons per every displaced spin.
 Then, in magnon description, a commensurate-incommensurate transition involves
a condensation of an infinite number of magnons.
One can imagine, by analogy with coupled superconducting and spin density orders \cite{sachdev2001},
that proliferation of SC domain walls,
depicted in Fig.~\ref{fig:SC},
may cause the appearance of an incommensurate modulation of
  $\langle S^z\rangle$
due  to ``density-density'' type coupling between
the magnon density  and the density of domain walls.
Whether or not this is the case requires going beyond the instability condition \eqref{mon_3}
and analyzing excitation spectrum and inter-pair interactions within the spin-current phase~\cite{noz}.

{\underline{{\em Conclusions.}}~~~~We have described a novel two-magnon pairing instability of the
 up-up-down phase of the spatially anisotropic
triangular lattice antiferromagnet in a magnetic field.
 The magnon pairing is of ``inter-band" type
 in that the condensate is made out of bosons from the two different spin-wave branches.
This instability pre-empts a single-magnon condensation for arbitrary spin $S$ and gives rise to a
 highly unconventional 2D order in which transverse spin components are
disordered, yet the ground state has a non-zero vector chirality on every lattice bond and circulating spin currents in every elementary triangle.
This state breaks $Z_2$ chiral symmetry but preserves  $U(1)$ symmetry of rotations about the field direction.
The development of such a phase can be thought of as a spontaneous generation of the \DM~ interaction.
This new state
  exhibits a  magneto-electric effect, which gives rise to a non-trivial linear dependence of spin-wave excitations on the applied electric field
  $E$, and also has staggered magnetic moments, which can be measured by NMR and $\mu$SR.

We acknowledge illuminating discussions with L. Balents, C. Batista, A. Daley, L. Glazman, A. Furusaki, O. Kolezhuk, and O. Sushkov.
%OS aug7
We thank Qi Hu for pointing out  inconsistencies in Eq.(9) and in several related formulas in the supplementary part of the paper.
This work was supported by  DOE DE-FG02-ER46900  (A.V.Ch.) and by NSF DMR-1206774 (O.A.S.).

%\newpage
\section{Supplementary material for ``Spin-current order in anisotropic triangular antiferromagnets'' by A. V. Chubukov and O. A. Starykh}
\label{sec:suppl}
\setcounter{page}{1}
\setcounter{equation}{0}
\setcounter{figure}{0}
\subsection{ One-magnon excitations in  the UUD phase}

One-magnon excitations in the UUD phase in the anisotropic case ($J' < J$) have been analyzed in Ref. \onlinecite{alicea_a}.
For completeness, we present here the details of the derivation. We will use some of intermediate formulas in the next section,
when we derive the pairing interaction between magnons.

Spin-wave description of the UUD state proceeds as follows.
First, we use a three-sublattice
representation where spins point up on sublattices A and B and down on sublattice C, and introduce Holstein-Primakoff bosons $a$, $b$, and $c$ respectively.
Spins on the A
 sublattice
 are described by
\bea
&&S^z_A({\bf r}) = S - a^\dagger_{\bf r} a_{\bf r}, \nonumber \\
&& S^+_A({\bf r}) = \sqrt{2S} \sqrt{1- \frac{a^\dagger_{\bf r} a_{\bf r}}{2S}} a_{\bf r} \approx \sqrt{2S}
(1 - \frac{a^\dagger_{\bf r} a_{\bf r}}{4S}) a_{\bf r},
\label{ap:Sa}
\eea
and spins on the B sublattice are represented by the same expressions with $a_{\bf r}$ replaced by $b_{\bf r}$. The
 expansion of the square-root is valid for $S \gg1$, and below we assume that $S$ is indeed large.
 Spins on C sublattice points opposite to those on A and B sublattices, and
we have
 \be
S^z_C({\bf r}) = -S + c^\dagger_{\bf r} c_{\bf r} , S^-_C({\bf r}) = \sqrt{2S} (1 - \frac{c^\dagger_{\bf r} c_{\bf r}}{4S}) c_{\bf r}.
\label{ap:Sc}
\ee
Plugging this in \eqref{1}, we obtain spin-wave Hamiltonian as the sum of the linear (harmonic)
 term
 $H^{(2)}$ and the interaction terms $H^{(4)}$
\begin{eqnarray}
H^{(2)} &=& S \sum_{\bf k} \Big[[\gamma_{\bf k} a^\dagger_{\bf k} b_{\bf k}
  + \gamma_{{\bf k}}(b^\dagger_{\bf k} c^\dagger_{-{\bf k}}+ c^\dagger_{{\bf k}}a^\dagger_{-\bf k} )  + \text
{h.c.}]
\nonumber\\
&& \!\!\!\!\!\!\!\!\!\! +
h(a^\dagger_{\bf k} a_{\bf k} + b^\dagger_{\bf k} b_{\bf k}) + (2h_0-h) c^\dagger_{\bf k} c_{\bf k}\Big].
  \label{eq:H2}
\end{eqnarray}
Here
 the sum extends over the magnetic Brillouin zone, whose area is $1/3$ of the total area of the Brillouin zone,
 $\gamma_{\bf k} = J e^{i k_x} + 2J' \cos(\sqrt{3}k_y/2) e^{-ik_x/2}$ and $h_0 = J+2J'$.
The interaction term $H^{(4)} = H^{(4)}_z + H^{(4)}_\perp$ is the sum of the transverse ($\perp$) and longitudinal ($z$)
  contributions
  %OS aug2
\begin{eqnarray}
&&H^{(4)}_\perp =(-\frac{J}{4}) \frac{3}{N}\sum_{{\bf k}_1 - {\bf k}_3} \{ \gamma_1 (c_1^\dagger a_2^\dagger a_3^\dagger a_{1+2+3} +
b_1^\dagger c_2^\dagger c_3^\dagger c_{1+2+3} \nonumber\\
&&+ a_1^\dagger b_2^\dagger b_3 b_{1+2-3}) + \gamma_{-1}(b_1^\dagger a_2^\dagger a_3 a_{1+2-3} + a_1^\dagger c_2^\dagger c_3^\dagger c_{1+2+3} \nonumber\\
&& + c_1^\dagger b_2^\dagger b_3^\dagger b_{1+2+3}) + \text{h.c.}\},
\label{eq:H4perp}
\end{eqnarray}
 and
\begin{eqnarray}
&&H^{(4)}_z =\frac{3J}{N}\sum_{{\bf k}_1 - {\bf k}_3} \{ \gamma_{1-2} a_1^\dagger a_2 b_3^\dagger b_{1-2+3} +
\nonumber\\
&&- \gamma_{1-2} b_1^\dagger b_2 c_3^\dagger c_{1-2+3} - \gamma_{2-1} a_1^\dagger a_2 c_3^\dagger c_{1-2+3}\}.
\label{eq:H4z}
\end{eqnarray}
 Here we denote
 for brevity $\gamma_1 \equiv \gamma_{\bf k_1}, a_1 \equiv a_{\bf k_1}$ and so forth.

 If we take only the quadratic part \eqref{eq:H2}  and
diagonalize it, we find that the UUD phase is stable for just one field value $h = 3J$ for the isotropic case ($J=J'$) and is
 unstable for all fields when $\delta \neq 0$. (In the notations which we used in the text, the degree of anisotropy is measure in terms
 of dimensionless parameter  $\delta = (40S/3) (1 - J'/J)^2$, Ref.\onlinecite{alicea_a}).
However, interactions between magnons stabilize the UUD phase over the
finite $\delta$ range.
To see this, we first modify the quadratic form by adding the leading $1/S$ Hartree-type self-energy corrections from
 4-boson interaction terms  \eqref{eq:H4perp} and \eqref{eq:H4z},
 and then diagonalize the effective quadratic Hamiltonian. Since we are interested in large $S$ and small anisotropies (when $\delta = O(1)$) and
  in field range near $h = 3 J$, Hartree corrections can be computed at $\delta =0$ and $h = 3J$, when classical UUD state is critical and its spin-wave excitation spectrum
   does not contain complex modes.
In 2D Hartree corrections are all finite, and adding them to \eqref{eq:H2} we
obtain  the harmonic Hamiltonian of the UUD state in the form
 \begin{eqnarray}
H_{\text{uud}} &=& S \sum_{\bf k} \Big[[{\tilde \gamma}_{1,\bf k} a^\dagger_{\bf k} b_{\bf k}
  + {\tilde \gamma}_{2,{\bf k}}(b^\dagger_{\bf k} c^\dagger_{-{\bf k}}+ c^\dagger_{{\bf k}}a^\dagger_{-\bf k} )  + \text
{h.c.}]
\nonumber\\
&& \!\!\!\!\!\!\!\!\!\! + (h + \Sigma_1)
(a^\dagger_{\bf k} a_{\bf k} + b^\dagger_{\bf k} b_{\bf k}) + (2h_0+\Sigma_2-h) c^\dagger_{\bf k} c_{\bf k}\Big].
  \label{eq:uud1}
\end{eqnarray}
 where
 \be
 {\tilde \gamma}_{j,{\bf k}} = \gamma_{\bf k} + \Sigma^\prime_j
 \ee
and the self-energy components are $\Sigma_1 = 0.14J/S$, $\Sigma_2 = 0.67J/S$,
$\Sigma'_{1,{\bf 0}} = -0.11J/S$ and $\Sigma'_{2,{\bf 0}} = 0.18J/S$.
Observe that \eqref{eq:uud1} reduces to \eqref{eq:H2} in the $S\to\infty$ limit.

Alternatively, one could first diagonalize the linear spin-wave part \eqref{eq:H2}, express the 4-boson
interaction part in terms of new operators of system's eigen-modes  and then correct magnon dispersion by adding
to it quadratic terms in new operators,
  which appear as a result of normal-ordering  of Eqs. \eqref{eq:H4perp} and \eqref{eq:H4z} in the new basis.
  Such a procedure was first applied by Oguchi and is known as  Oguchi's corrections~\cite{oguchi}

Low-energy excitations near ${\bf k} = 0$ encode the important physics, and in
this region analysis of Eq.\ \eqref{eq:uud1} simplifies considerably.  Here we have
${\tilde \gamma}_{j,{\bf k}} \approx {\bar \gamma}_{j,{\bf k}} + i \Gamma_{\bf k}$, with
\begin{equation}
{\bar \gamma}_{j,\bf k} = h_0 + \Sigma'_{j,{\bf 0}} - \frac{3}{4} J k^2,~~\Gamma_{\bf k} =  (J-J') k_x.
\label{eq:uud2}
\end{equation}

Diagonalization of Eq.\ (\ref{eq:uud1})
 proceeds in two steps. We first diagonalize $H_{\text{uud}}'$, which is obtained
from \eqref{eq:uud1} by setting $\Gamma_{\bf k} = 0$. This is done by introducing new operators
\be
p_k = (a_k - b_k)/\sqrt{2}, d_k = (a_k + b_k)/\sqrt{2}
\label{ap:p-d}
\ee
which decouple in
 $H_{\text{uud}}'$. However, $d$-mode couples to $c$-boson
 via ${\bar \gamma}_{j,\bf k} (d_k^\dagger c_{-k}^\dagger + \text{h.c.})$ term, and to diagonalize the full quadratic Hamiltonian one needs
  to apply rotation
\bea
d_k &=& \cosh\theta_k u_k + \sinh\theta_k v_{-k}^\dagger, \nonumber\\
c_{-k}^\dagger &=& \sinh\theta_k u_k + \cosh\theta_k v_{-k}^\dagger ,
\label{ap:d-c}
\eea
where
 \be
\tanh(2 \theta_{\bf k}) = \frac{-2\sqrt{2}{\bar\gamma}_{2,{\bf k}}}{2h_0+\Sigma_1+\Sigma_2+{\bar\gamma}_{1,{\bf k}}} \to -\frac{2\sqrt{2}}{3}.
\label{ap:2}
\ee
The quadratic Hamiltonian in terms of $u_k$, $v_k$ and $p_k$ operators is
\begin{equation}
H_{\text{uud}}' = S\sum_{\bf k}[ \omega_p p^\dagger_{\bf k} p_{\bf k}
+ \omega_v v^\dagger_{\bf k} v_{\bf k}
+ \omega_u u^\dagger_{\bf k} u_{\bf k}].
\label{eq:H-uud-prime}
\end{equation}
The $u$ boson describes the precession of the total magnetization. The
 corresponding frequency
  $\omega_u({\bf 0}) \sim h$ is large,
which implies that this mode is irrelevant for low-energy physics.
The two remaining bosons, $p$ and $v$,
are the low-energy modes of interest.  For small ${\bf k}$
 \begin{eqnarray}
\omega_p({\bf k}) &=& [h-h_{c1}^0+2(J-J')] + \frac{3}{4}J k^2, \nonumber\\
\omega_v({\bf k}) &\approx& [h_{c2}^0-2(J-J')-h] + \frac{9}{4}J k^2.
\label{eq:uud3}
\end{eqnarray}
 where $h_{c1}^0 = 3J - 0.5 J/(2S)$ and $h_{c2}^0 = 3J + 1.3 J/(2S)$ are the boundaries of the UUD phase in the isotropic case.

The next step is to account for the
  remaining part of $H_{\rm uud}$ in \eqref{eq:uud1}, which is proportional to $\Gamma_{\bf k} =  (J-J') k_x$.
 The corresponding term, which
 we denote by $H_{\text{uud}}''$,
  has the form
 \begin{equation}
H_{\text{uud}}'' = 3 iS\sum_{\bf k} \Gamma_{\bf k}
(p_{\bf k} v_{-{\bf k}} - \text{h.c.}) ,
\label{eq:uud4}
\end{equation}

The diagonalization of
 $H'_{\rm uud} + H''_{\rm uud}$
 proceeds in the same way as before:
  we introduce new operators $d_{1,{\bf k}}$ and $d_{2,{\bf k}}$ as
 \bea
p_{\bf k} &=& \cosh\phi_{\bf k} d_{1,{\bf k}} + i \sinh\phi_{\bf k} d_{2,-{\bf k}}^\dagger,\nonumber\\
v_{-{\bf k}} &= & \cosh\phi_{\bf k} d_{2, -{\bf k}} + i \sinh\phi_{\bf k} d_{1,{\bf k}}^\dagger ,
\label{ap:p-v}
\eea
and choose $\phi_k$ to eliminate non-diagonal $d_{1,{\bf k}} d_{2,{-\bf k}}$ terms.
This last requirements leads to
\be
\tanh(2\phi_{\bf k}) = \frac{6(J-J')k_x}{\omega_p({\bf k}) + \omega_v({\bf k})} = \frac{6(J-J')k_x}{\Delta h + 3J (k_x^2 + k_y^2)}.
\label{ap:tanh}
\ee
Here $\Delta h = h_{c2}^0 - h_{c1}^0 = 1.8J/(2S)$ is the width of the UUD phase
 at $J'=J$.
The diagonalized quadratic Hamiltonian is, up to a constant,
\begin{equation}
  H_{\text{uud}} = S\sum_{\bf k}[ \omega_1 d^\dagger_{1,\bf k}d_{1,\bf k}
+ \omega_2 d^\dagger_{2,\bf k} d_{2,\bf k}],
\label{ap:H-uud-prime}
\end{equation}
where at small ${\bf k}$
\begin{eqnarray}
 \omega_{1,2}({\bf k}) &=& \pm \frac{1}{2}(\omega_p - \omega_v) + \frac{1}{2} \sqrt{(\omega_p + \omega_v)^2 - 36 (J-J')^2 k_x^2}\nonumber\\
 &=&
 \pm \left(h-h_0-\frac{1}{5S}J-\frac{3}{4}J{\bf k}^2\right) + \frac{3J Z_{\bf k}}{20S} ,
\label{ap:uud5}
\end{eqnarray}
with $Z_{\bf k} =\sqrt{9 + 10 S (6 {\bf k}^2 -3\delta k_x^2 + 10 S {\bf k}^4)}$.

 The UUD phase is stable with respect to small perturbations when both modes are positive.
 The full analysis  has been done in Ref. \onlinecite{alicea_a}, where it was shown that UUD phase survives up to
 $\delta =4$.
  For our purposes, we focus on the region near the end point.
  The UUD phase is stable at  $h_{c2} > h > h_{c1}$, where
  %OS aug2
  \beq
 h_{c1} = h_{\rm end} - \frac{27J}{40S} \sqrt{\frac{4-\delta}{3}},
 h_{c2} = h_{\rm end} + \frac{27J}{40S} \sqrt{\frac{4-\delta}{3}},
\label{hc12}
\eeq
and $h_{\rm end} = h_0 (1 + 17/120S)$.
Near the lower critical field $h_{c1}$, the mode $\omega_1 (k)$ softens at $\pm {\bf k}_{1} = (k_1,0)$, where
$k_1 \approx (3/(10S))^{1/2} (1 + \sqrt{(4-\delta)/12})$. Near the upper critical field $h_{c2}$, the mode $\omega_2 (k)$ softens at $\pm {\bf k}_{2} = (k_2,0)$, where
$k_2\approx (3/(10S))^{1/2} (1- \sqrt{(4-\delta)/12})$.
At $\delta =4$, the two critical fields become equal  $h_{c1} = h_{c2} = h_{\rm end}$,
 and  both modes soften at the same $k^2_1 = k^2_2 = k^2_0 = 3/(10S)$.
At this $\delta$, the excitation spectra are
\bea
&&\omega_1 (k) = \frac{3J}{2} \sqrt{(k^2_x-k^2_0)^2 + 4 k^2_0 k^2_y} - \frac{3J}{4} (k^2_x - k^2_0) \nonumber \\
&& \omega_2 (k) = \frac{3J}{2} \sqrt{(k^2_x-k^2_0)^2 + 4 k^2_0 k^2_y} + \frac{3J}{4} (k^2_x - k^2_0)
\label{s_5}
\eea
Observe that  at this point 
%OS aug2
$6 (J-J')k_0 = 18J/(10S) = 2 \Delta h$ and $3J k^2_0 = \Delta h$, hence
 $\tanh(2\phi_{k_0,0}) = 1$ in (\ref{ap:tanh}), i.e.,
 $\phi_{\bf k}$ diverges at $\pm {\bf k}_0 = (k_0,0)$. The divergence of $\phi$ implies that the coherence factors
$\cosh \phi_k$ and $\sinh \phi_k$ strongly diverge too.
At small deviations from $\pm {\bf k}_0$ and $\delta =4$,
\bea
&& \cosh2\phi_{\bf k} \approx \sinh 2\phi_{\bf k} = \frac{1}{2}e^{2\phi_{\bf k}}
\nonumber\\
&&=\frac{2k^2_0}{\sqrt{(k^2_x- k^2_0)^2 + 4 k^2_0 k_y^2 + (4-\delta)k_0^4}} = f^2 (k)
\label{s_1}
\eea

Below we will need to express bosons $a,b,c$ via the low-energy
eigen-modes $d_1$ and $d_2$.
Working backward through transformations \eqref{ap:p-d}, \eqref{ap:d-c}, and \eqref{ap:p-v}, we obtain
\begin{eqnarray}
a_{\bf k} &= &\frac{1}{\sqrt{2}} \{ (\cosh \phi_{\bf k} + i \sinh\phi_{\bf k}) d_{1,{\bf k}} - \nonumber\\
&&-(\cosh \phi_{\bf k} - i \sinh\phi_{\bf k}) d_{2,-{\bf k}}^\dagger\}, \nonumber\\
b_{\bf k} &= &\frac{-1}{\sqrt{2}} \{ (\cosh \phi_{\bf k} - i \sinh\phi_{\bf k}) d_{1,{\bf k}} + \nonumber\\
&&+(\cosh \phi_{\bf k} + i \sinh\phi_{\bf k}) d_{2,-{\bf k}}^\dagger\}, \nonumber\\
c_{\bf k} &= &\sqrt{2} \{ \cosh \phi_{\bf k}  d_{2,{\bf k}} - i \sinh\phi_{\bf k} d_{1,-{\bf k}}^\dagger\}.
\label{ap:abc}
\end{eqnarray}
Near ${\bf k} \pm {\bf k}_0$ and $\delta =4$, $\phi_{\bf k}$ is large, and using
(\ref{s_1}) one can simplify the  transformation to
\begin{eqnarray}
&&a_{\bf k} = \frac{f({\bf k})}{\sqrt{2}}(e^{i s_{\bf k}} d_{1,{\bf k}} - e^{- i s_{\bf k}} d_{2, -{\bf k}}^\dagger),\nonumber\\
&&b_{\bf k} = -\frac{f({\bf k})}{\sqrt{2}}(e^{-i s_{\bf k}} d_{1,{\bf k}} + e^{ i s_{\bf k}} d_{2, -{\bf k}}^\dagger),\nonumber\\
&&c_{\bf k} = f({\bf k})(d_{2,{\bf k}} - e^{ i 2s_{\bf k}} d_{1, -{\bf k}}^\dagger).
\label{eq:abc_1}
\end{eqnarray}
Here $s_{\bf k} = \pi \text{sign}(k_x)/4$.

\subsection{Derivation of the pairing interaction between $d_1$ and $d_2$ magnons}

To obtain the interaction between low-energy magnons, one has to express the interaction Hamiltonian
$H^{(4)} = H^{(4)}_z + H^{(4)}_\perp$ written in terms of $a, b,$ and $c$ bosons, Eqs. \eqref{eq:H4perp} and \eqref{eq:H4z},
 via $d_1$ and $d_2$ operators with the help of  Eq. \eqref{ap:abc}, and find which of the generated interaction terms are the strongest.
 This procedure is straightforward but time-consuming. We analyzed pairing interaction with zero total momentum of the
 pair and with total momentum $2k_0$.  We found that the interaction matrix elements are much stronger for the former case (zero total momentum pairs).
   The computational procedure is similar in both cases and we present only the
 details of the derivation of the strongest interaction.

 Because $k_0 = (3/(10S))^{1/2}$  is small, we approximate the factors $\gamma_k$ in   Eqs. \eqref{eq:H4perp} and \eqref{eq:H4z}
 by  their values at $k=0$, i.e., approximate $\gamma_k$ by $\gamma_0 = J + 2J' \approx 3J$.
We verified that that keeping the momentum dependence of $\gamma_{\bf k}$'s only gives rise to
 irrelevant small corrections.

 We assume and then verify that the dominant contribution to magnon pairing comes from
  momenta near $\pm k_0$. To obtain the pairing vertices with zero total momentum,
  it is then convenient to introduce pair operators $\Psi_R({\bf q}) = d_{1,{\bf k}_0 + {\bf q}} d_{2,-{\bf k}_0 - {\bf q}}$ and
$\Psi_L({\bf q}) = d_{1,-{\bf k}_0 + {\bf q}} d_{2,{\bf k}_0 - {\bf q}}$, where $|{\bf q}| << k_0$.
 Expressing $H^{(4)}$ in terms of $d_1$ and $d_2$
we find after long but straightforward calculation that the pairing vertex can be expressed as
\begin{eqnarray}
&&{\cal H}_{d_1d_2}^{(4)} = -\frac{3J}{2} \frac{3}{N} \sum_{\bf q, p} \Big( (1+ \cosh2\phi_q \cosh2\phi_p + \nonumber\\
&&+\sinh2\phi_q \sinh2\phi_p)\times (\Psi_R^\dagger(q) \Psi_L(p) + \Psi_L^\dagger(q) \Psi_R(p))
\nonumber\\
&&+ (1+ \cosh2\phi_q \cosh2\phi_p - \sinh2\phi_q \sinh2\phi_p) \nonumber\\
&&\times(\Psi_R^\dagger(q) \Psi_R(p) + \Psi_L^\dagger(q) \Psi_L(p))+\nonumber\\
&& + (-1+ \cosh2\phi_q \cosh2\phi_p + \sinh2\phi_q \sinh2\phi_p) \times \nonumber\\
&&\times(\Psi_R^\dagger(q) \Psi_L^\dagger(p) + \text{h.c.})
\nonumber\\
&& + (-1+ \cosh2\phi_q \cosh2\phi_p - \sinh2\phi_q \sinh2\phi_p) \times \nonumber\\
&&\times(\Psi_R^\dagger(q) \Psi_R^\dagger(p) + \Psi_L^\dagger(q) \Psi_L^\dagger(p)+ \text{h.c.})\Big).
\label{ap:H4long}
\end{eqnarray}

We see that there are two types of pairing vertices: the ones with transferred momentum (for a given boson kind) of the order $2k_0$
(these are $\Psi_L \Psi_R$ terms), and
 the ones with transferred momentum near zero  ($\Psi_L \Psi_L$ and $\Psi_R \Psi_R$ terms).
 For the first set of terms, the vertex contains
 $\cosh2\phi_q \cosh2\phi_p + \sinh2\phi_q \sinh2\phi_p$ and diverges at $\delta =4$ in the limit ${\bf p}, {\bf q} \to 0$.
 For the second set,
  the vertex contains $\cosh2\phi_q \cosh2\phi_p - \sinh2\phi_q \sinh2\phi_p$, and the leading divergent terms cancel out.
  As a result, the vertex with momentum transfer near $2k_0$ is much stronger.
  Keeping only this vertex and using the asymptotic forms of $\cosh 2 \phi_{q/p}$ and $\sinh2\phi_{q/p}$ from Eq. (\ref{s_1}) we obtain
   \begin{eqnarray}
&&{\cal H}_{d_1d_2}^{(4)} = (3J) \frac{3}{N} \sum_{\bf q, p} f^2 (q) (\Psi_R({\bf q}) - \Psi_R^\dagger({\bf q}))\nonumber\\
&&\times f^2(p) (\Psi_L({\bf p}) - \Psi_L^\dagger({\bf p})).
\label{ap:H4short}
\end{eqnarray}

\subsection{Solution of the gap equation}

As is customary in superconductivity studies, we add to the Hamiltonian  infinitesimally small pairing terms  $Q_L = \Phi_{0,L} (q) \Psi_L (q),
 Q_R = \Phi_{0,R} (q) \Psi_L (q)$ with generally complex $\Phi_{0,L}$ and $\Phi_{0,R}$ and obtain the renormalized
 $\Phi_L (q)$  and $\Phi_R (q)$ by summing up ladder series of vertex corrections. At the pairing instability, the pairing susceptibility diverges, and the
 equations for $\Phi_L (q)$  and $\Phi_R (q)$ have solutions even when we set bare  $\Phi_{0,L} (q)$ and $\Phi_{0,R} (q)$ to zero.

 The diagrams for the fully renormalized  $\Phi_L (q)$  and $\Phi_R (q)$ at the instability are shown in Fig.~\ref{fig:diag}.
 One can easily make sure that the full set of coupled equations for different $\Phi$ and $\Phi^*$ separates into two independent sets
 for $\Phi_L$ and $\Phi^*_R$ and for
$\Phi_R$ and $\Phi^*_L$.  Because the momentum dependence of the pairing vertex in
Eq. (\ref{ap:H4short}) is factorized into
$f^2 (p)  f^2 (q)$, we search for the solution in the form $\Phi_L (q) = \tilde\Upsilon f^2 (q)$.
Substituting this form into the diagrams and using one-magnon dispersions from (\ref{s_5}) we obtain after a simple algebra
\bea
\tilde\Upsilon &=& (\tilde\Upsilon -{\tilde\Upsilon}^*) \frac{3}{NS} \sum_p \frac{3J f^4 (p) }{\omega_1 (k_0 +p) + \omega_2 (k_0 +p)} \nonumber \\
&&= i ({\rm Im} \tilde\Upsilon) \frac{3}{NS} \sum_{\tilde p} \frac{k_0}{(p^2 + (1-\delta/4) k^2_0)^{3/2}}
\label{s_6}
\eea
We used the fact that $\omega_1 (k_0 +p) + \omega_2 (k_0 +p) = 6J k^2_0/f^2 ({\bf p})$ (see Eq. (\ref{s_5})).
It is obvious from Eq. (\ref{s_6}) that $\tilde\Upsilon$ should be purely imaginary, $\tilde\Upsilon = i \Upsilon$.
Substituting this into (\ref{s_6}), we find that the equation for $\Upsilon$ has a non-trivial solution when
 \be
 1 = \frac{1}{S} \frac{3}{N} \sum_p \frac{k_0}{|{\bf p}|^3}
 \label{s_7}
 \ee
 Generalizing this to $\delta \leq 4$ case, and replacing the sum over $p$ by integral we find that the condition on the pairing instability reduces to
 \be
 \int \frac{d^2 p}{(p^2 + (1-\delta/4)k^2_0)^{3/2}} = \tilde{a} \frac{S}{k_0}
 \ee
  where $\tilde{a} = O(1)$.
  Evaluating the integral we find that the two-magnon instability occurs at $4-\delta = O(1/S^2)$.

   \begin{figure}
\begin{center}
  \scalebox{0.8}{\includegraphics[width=\columnwidth]{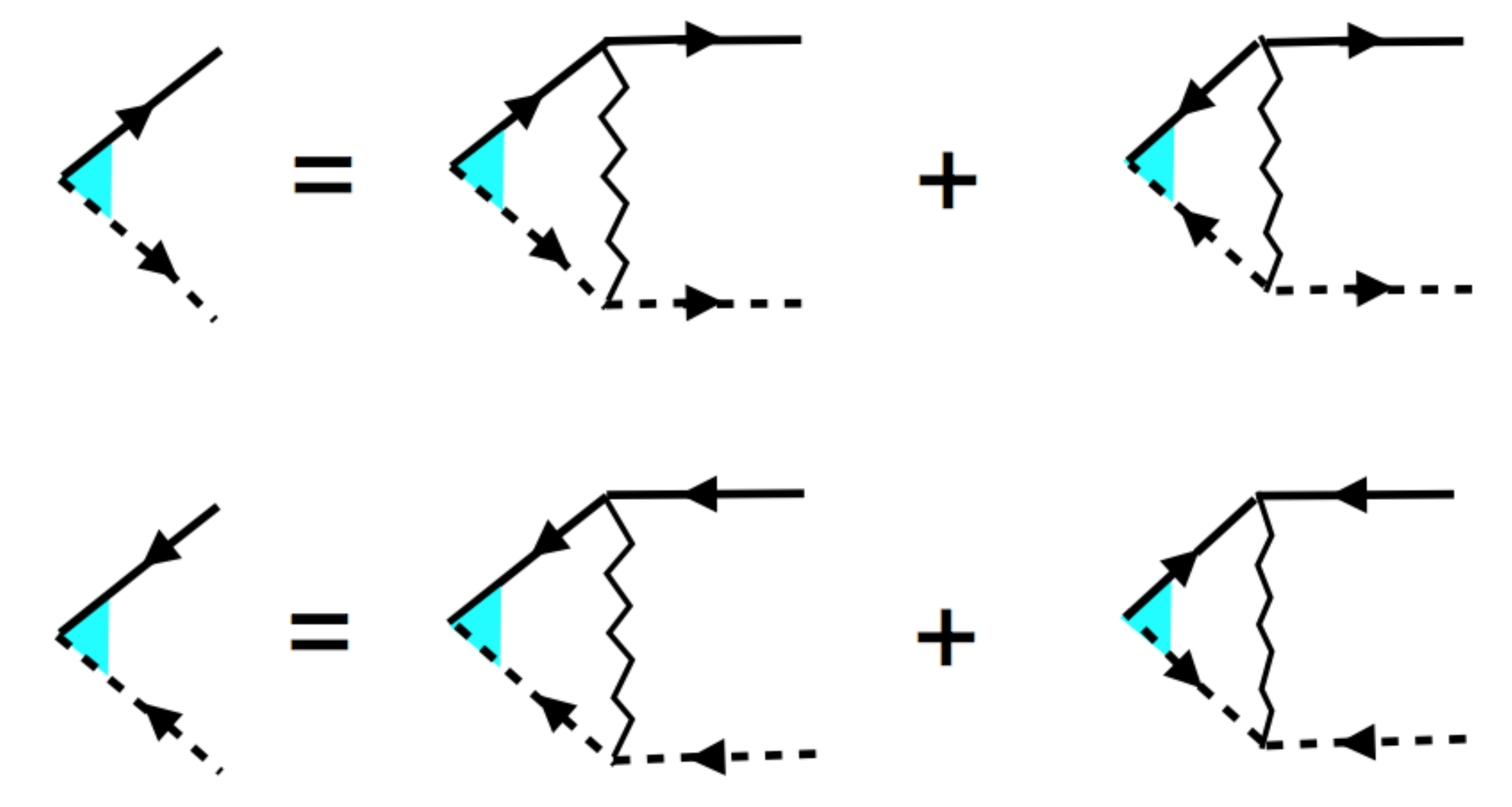}}
  \end{center}
\caption{(Color online) Coupled set of diagrams for anomalous vertices $\Phi_L$ (upper line)  and $\Phi_R^* $ (lower line).}
\label{fig:diag}
\end{figure}

\subsubsection{Alternative derivation of Eq. (\ref{s_7}).}

 We start from simple observation that
\be
[{\cal H}^{(2)}_{\text{uud}}, \Psi_R({\bf q})] = - S\Big(\omega_1({\bf k}_0 + {\bf q}) + \omega_2(-{\bf k}_0 - {\bf q})\Big)  \Psi_R({\bf q}),
\ee
and similarly for $ \Psi_L({\bf q})$.
 We also see that
\be
[ \Psi_R({\bf q}),  \Psi_R^\dagger({\bf p})] = \delta_{q,p}(1 + d^\dagger_{1,{\bf k}_0 + {\bf q}} d_{1,{\bf k}_0 + {\bf q}} + d^\dagger_{2,{\bf k}_0 - {\bf q}} d_{2,{\bf k}_0 - {\bf q}})
\label{ap:comm}
\ee
and $[ \Psi_R({\bf q}),  \Psi_L^\dagger({\bf p})] = 0$ because for $q,p \ll k_0$, the two pairs do not overlap in momentum space.
 Inside the UUD plateau $\langle d^\dagger_{1/2,{\bf k}} d_{1/2,{\bf k}}\rangle = 0$, where the average is over the ground state.
Hence the right-hand side of \eqref{ap:comm} can be replaced by $\delta_{q,p}$, implying canonical bosonic commutation relations for pairs $\Psi_{R/L}$.
This allows for an easy derivation of the equations of motion for pair operators. We obtain
\bea
i\partial_t \Psi_R({\bf k}) &=& \Omega_{\bf k} \Psi_R({\bf k}) - \frac{3}{N} \sum_{\bf p} \Phi({\bf p}, {\bf k}) (\Psi^\dagger_L({\bf p}) - \Psi_L({\bf p})),\nonumber\\
i\partial_t \Psi_L({\bf k}) &=& \Omega_{\bf k} \Psi_L({\bf k}) - \frac{3}{N} \sum_{\bf p} \Phi({\bf p}, {\bf k}) (\Psi^\dagger_R({\bf p}) - \Psi_R({\bf p})),\nonumber\\
i\partial_t \Psi_R^\dagger({\bf k}) &=& -\Omega_{\bf k} \Psi_R^\dagger({\bf k}) - \frac{3}{N} \sum_{\bf p} \Phi({\bf p}, {\bf k}) (\Psi^\dagger_L({\bf p}) - \Psi_L({\bf p})),\nonumber\\
i\partial_t \Psi_L^\dagger({\bf k}) &=& -\Omega_{\bf k} \Psi_L^\dagger({\bf k}) - \frac{3}{N} \sum_{\bf p} \Phi({\bf p}, {\bf k}) (\Psi^\dagger_R({\bf p}) - \Psi_R({\bf p})), \nonumber \\
&&
\eea
where $\Omega_{\bf k} = S (\omega_1({\bf k}_0 + {\bf q}) + \omega_2(-{\bf k}_0 - {\bf q}))$
 and $\Phi (q,p) \approx -3J f^2 (p) f^2 (q)$.
We now Fourier transform $t$-dependence
 (
 $i\partial_t \to \omega$)
  and
 set $\omega = 0$
  because we are seeking the condition for the pair condensation.
  We then take expectation values of both sides of equations and form appropriate linear combinations to obtain
\bea
\langle \Psi_R^\dagger({\bf k}) - \Psi_R({\bf k})\rangle &=& \frac{
 6J
 f^2({\bf k})}{\Omega_{\bf k}} \frac{3}{N}\sum_{\bf p} f^2({\bf p})
\langle \Psi_L^\dagger({\bf p}) - \Psi_L({\bf p})\rangle\nonumber\\
\langle \Psi_L^\dagger({\bf k}) - \Psi_L({\bf k})\rangle &=& \frac{
6J
f^2({\bf k})}{\Omega_{\bf k}} \frac{3}{N}\sum_{\bf p} f^2({\bf p})
\langle \Psi_R^\dagger({\bf p}) - \Psi_R({\bf p})\rangle. \nonumber \\
&&
\label{s_8}
\eea
where we used $\gamma_k \approx 3J$.
We immediately see that $\Phi_R$ and $\Phi_L$ must be purely imaginary, and $\Phi_{L,R} (k) \propto f^2 (k)$.
The self-consistency condition then gives
\be
\frac{3}{N}\sum_{\bf p} \frac{6J f^4({\bf p})}{\Omega_{\bf p}} = \frac{1}{S} \frac{3}{N} \sum_p \frac{k_0}{|{\bf p}|^3}
 = 1,
\ee
which is  the same condition as Eq. (\ref{s_7}).

\subsection{Pair condensation and spontaneous generation of  Dzyaloshinskii-Moria interaction}

 We first observe that Eq. \eqref{ap:H4short} can be re-written as
\be
{\cal H}_{d_1d_2}^{(4)} =  -(J+2J') \frac{3}{N} {\cal H}^{\rm DM}_{+k_0} ~{\cal H}^{\rm DM}_{-k_0}.
\label{ap:dm2}
\ee
where
\bea
&&{\cal H}^{\rm DM}_{k_0} = i \sum_{{\bf q}} f^2({\bf q}) \left(\Psi_{R}({\bf q}) - \Psi_{R}^\dagger({\bf q})\right), \nonumber \\
&& {\cal H}^{\rm DM}_{-k_0} = i \sum_{{\bf q}} f^2({\bf q}) \left(\Psi_{L}({\bf q}) - \Psi_{L}^\dagger({\bf q})\right).
\label{ap:Hdm}
\eea
Note that the integrands in (\ref{ap:Hdm}) is the same as the
two-magnon order parameter
 \be
 i \sum_{{\bf q}\in \pm {\bf k}_0} f^2({\bf q}) \left(d_{1,{\bf q}} d_{2,-{\bf q}} - d^\dagger_{1,{\bf q}} d^\dagger_{2,-{\bf q}}\right).
\ee
Hence, once two-magnon condensation occurs, $\langle{\cal H}^{\rm DM}_{\pm k_0}\rangle$  acquires a non-zero  expectation value,
proportional to $\Upsilon$, i.e., the Hamiltonian acquires an extra term
\be
{\cal H}_{d_1d_2}^{(4)} =  D \left[{\cal H}^{\rm DM}_{+k_0} + {\cal H}^{\rm DM}_{-k_0}\right].
\label{s_9}
\ee
where
%OS aug2
\be
D = -(J+2J') \frac{3}{N} \langle{\cal H}^{\rm DM}_{+k_0}\rangle = (J+2J') \Upsilon .
\label{s_10}
\ee
(See next section for a very similar calculation.)
We now compare the Hamiltonian in Eq. (\ref{s_9}) with the one which describes Dzyaloshinskii-Moria (DM) interaction in a triangular magnet
and show that they are identical.
The  DM interaction on a triangular lattice reads \cite{alicea_a2}
\bea
H_{\rm DM-latt} &=& \hat{z}\cdot\sum_{\bf r} \Big( {\bf S}_C({\bf r})\times {\bf S}_B({\bf r}) + {\bf S}_A({\bf r})\times {\bf S}_C({\bf r}) + \nonumber\\
&& + {\bf S}_B({\bf r})\times {\bf S}_C({\bf r})\Big).
\label{ap:dm-latt}
\eea
 In our  case  $z$ coincides with the direction of external field.
 Expressing the spins in terms of Holstein-Primakoff bosons $a, b$, and $c$
 and transforming to $d_1$ and $d_2$ operators using (\ref{eq:abc_1}) we obtain that after some algebra
\eqref{ap:abc} gives
\be
H_{\rm DM-latt} = 6S ({\cal H}^{\rm DM}_{+k_0}  + {\cal H}^{\rm DM}_{-k_0}).
\ee
Comparing with (\ref{s_9}) we immediately see that the appearance of a  two-magnon condensate with an imaginary amplitude $\Upsilon$
can be viewed as a spontaneous generation of DM interaction with the coupling $D \propto \Upsilon$.

\subsection{Structure of spin currents}

The $z$-component of the spin current on the bond $\langle n,m\rangle$, connecting sites $n$ and $m$, is defined as
\be
J_{nm}^z = \frac{1}{2i} ( S^{-}_n S^+_m - S^+_n S^{-}_m).
\ee
 Re-expressing the r.h.s. of this expression in terms of $a$, $b$, and $c$ bosons, we find that spin currents along the bonds between spins from
 A, B, and C sublattices belonging to the same
 elementary triangle at a coordinate $r$  are determined by the following combinations
%AC added r
\bea
&&J^z_{CA} = i S (c^+_{\bf r} a^+_{\bf r} - c_{\bf r} a_{\bf r}), J^z_{CB} = i S (c^+_{\bf r} b^+_{\bf r} - c_{\bf r} b_{\bf r}), \nonumber\\
&&J^z_{AB} = i S (b^+_{\bf r} a_{\bf r} - a^+_{\bf r} b_{\bf r}).
\eea
 Using \eqref{eq:abc_1}, we obtain
\bea
\langle c_{\bf r} a_{\bf r} \rangle &=& \frac{3}{\sqrt{2} N} \sum_{k, q} e^{i {\bf r} \cdot ({\bf k} + {\bf q})} f({\bf k}) f({\bf q})
\langle (d_{2,{\bf k}} - e^{ i 2s_{\bf k}} d_{1, -{\bf k}}^\dagger)\nonumber\\
&&\times (e^{i s_{\bf q}} d_{1,{\bf q}} - e^{- i s_{\bf q}} d_{2, -{\bf q}}^\dagger)\rangle .
\eea
The condensate emerges at ${\bf q} = - {\bf k}$, and we obtain
%OS aug2
\bea
\label{aug2-1}
\langle c_{\bf r} a_{\bf r} \rangle &=& \frac{3}{\sqrt{2} N} \sum_{\bf k} f^2({\bf k}) e^{- i s_{\bf k}}
\langle d_{1, -{\bf k}} d_{2,{\bf k}} - d_{1, -{\bf k}}^\dagger d_{2,{\bf k}}^\dagger \rangle
\nonumber\\
&& =  \frac{3}{\sqrt{2} N} \sum_{\bf k} f^2({\bf k}) e^{- i s_{\bf k}} (2 i\Upsilon) 
\frac{3 J f^2({\bf k})}{S(\omega_{1} + \omega_{2})}\nonumber\\
&& = \frac{i\Upsilon}{S k_0^2} \frac{3}{ N} \sum_{\bf {\tilde k}} \Big(\frac{k_0^2}{{\tilde k}_x^2 + {\tilde k}_y^2 + (1-\delta/4) k_0^2}\Big)^{3/2} \nonumber\\
&& = i \Upsilon .
\eea
 Here ${\tilde k}$ is the deviation from  $\pm {\bf k}_0$,
 near which  the phase factor takes values $s_{\bf k} = \pm \pi/4$ correspondingly. 
%OS aug2
 The last line in the above equation is a direct consequence of the linearized ``gap" equation \eqref{mon_5} of the main text
 (which, of course, is the same as \eqref{s_6} and \eqref{s_7} of Supplementary Material), to which
 the right-hand-side of \eqref{aug2-1} reduces.
 
  Because $\langle c_{\bf r} a_{\bf r} \rangle$ is imaginary, it
   does not contribute to ${\bf S}_C \cdot {\bf S}_A \propto {\text{Re}}(S^+_C S^{-}_A)$, but the
spin current $J^z_{CA}$ becomes non-zero.
Similar calculation for other bonds shows that  $\langle c_{\bf r} a_{\bf r} \rangle = - \langle c_{\bf r} b_{\bf r} \rangle =\langle a_{\bf r}^\dagger b_{\bf r}\rangle$.
These relations fix the
 relative signs of spin currents and lead to two current patterns shown in Figure \ref{fig:SC} of the main text.

It is easy to generalize this calculation for the spins
 located at distance ${\bf R}$ apart  from each other
 (we assume that $R \gg k_0^{-1}$)
 %OS aug 2
\bea
&&\langle \hat{z}\cdot {\bf S}_C({\bf R}) \times {\bf S}_A(0)\rangle \propto S~ {\text{Im}} \langle c_{\bf R} a_0\rangle \nonumber\\
&&\sim \Upsilon k_0 \cos[{\bf k}_0 \cdot {\bf R}] \int_0^\infty d \tilde{k} \int_0^{2\pi} d\phi
\frac{\tilde{k} e^{i k R \cos[\phi]}}{(\tilde{k}^2 + (1-\delta/4)k_0^2)^{3/2}} \nonumber\\
&& \sim \Upsilon k_0 \cos[{\bf k}_0 \cdot {\bf R}] \int_0^\infty d \tilde{k} \frac{\tilde{k} J_0(\tilde{k} R)}{(\tilde{k}^2 + (1-\delta/4)k_0^2)^{3/2}} \nonumber\\
&& = \Upsilon k_0 \cos[{\bf k}_0 \cdot {\bf R}]  \xi e^{-R/\xi},
\eea
where $J_0$ is the Bessel function and $\xi^{-1} = k_0 \sqrt{1-\delta/4}$. The correlation decays exponentially for $R \gg \xi$.
 This implies that  $\xi$ has the meaning of the
  radius of a two-magnon bound state.  Using the relation $\delta_{\rm cr} =4 - O(1/S^2)$, we obtain  $\xi \sim S/k_0$.


\begin{references}
\bibitem{leon} L. Balents, Nature {\bf 464}, 199 (2010).
\bibitem{tokiwa06} Y. Tokiwa, T. Radu, R. Coldea, H. Wilhelm, Z. Tylczynski, and F. Steglich, \prb {\bf 73}, 134414 (2006).
\bibitem{takano08} N. Fortune, S. Hannahs, Y. Yoshida, T. E. Sherline, Y. Takano, T. Ono, and H. Tanaka, \prl {\bf 102}, 257201 (2009).
\bibitem{chubukov91} A. V. Chubukov and D. I. Golosov, J. Phys.: Condens. Matter {\bf 3}, 69 (1991).
\bibitem{misguich2001} G. Misguich, Th. Jolicoeur, and S. M. Girvin, \prl {\bf 87}, 097203 (2001).
\bibitem{alicea2007} J. Alicea and M. P. A. Fisher, \prb {\bf  75}, 144411 (2007).
\bibitem{alicea2} C. Griset, S. Head, J. Alicea, O. A. Starykh, \prb {\bf 84}, 245108 (2011).
\bibitem{alicea} J. Alicea, A. V. Chubukov, and O. A. Starykh, Phys. Rev. Lett. {\bf 102}, 137201 (2009).
\bibitem{zhit} T. Coletta, M. E. Zhitomirsky, and F. Mila, \prb {\bf 87}, 060407(R) (2013). % arXiv:1212.3086.
\bibitem{sdw2012} R. Chen, H. Ju, H. C. Jiang, O. A. Starykh, and L. Balents, \prb {\bf 87}, 165123 (2013). %arxiv:1211.1676 (2012).
\bibitem{suppl} See Supplementary Material for more details.
\bibitem{bal_nag} H. Katsura,  N. Nagaosa, and A. V. Balatsky,
Phys. Rev. Lett. 95, 057205 (2005); M. Mostovoy, Phys. Rev. Lett. 96, 067601 (2006).
\bibitem{batista_09} K. A. Al-Hassanieh, C. D. Batista, G. Ortiz, and L. N. Bulaevskii, Phys. Rev. Lett. {\bf 103}, 216402 (2009).
\bibitem{shura} A. A. Nersesyan, A. O. Gogolin, F. H. L. Essler, \prl {\bf 81}, 910 (1998).
\bibitem{kolezhuk} A. Kolezhuk and T. Vekua, \prb {\bf 72}, 094424 (2005).
\bibitem{akira-J1J2} T. Hikihara, T. Momoi, A. Furusaki, and H. Kawamura, \prb {\bf 81}, 224433 (2010).
\bibitem{chandra} P. Chandra, P. Coleman, and A.I. Larkin. J. Phys.: Condens. Matter {\bf 2}, 7933 (1990).
\bibitem{lauchli} A. L\"auchli, J.C. Domenge, C. Lhuillier, P. Sindzingre, and M. Troyer, \prl {\bf 95}, 137206 (2005).
\bibitem{andreev} A. F. Andreev and I. A. Grishchuk, Sov. Phys. JETP {\bf 60}, 267 (1984).
\bibitem{chubukov} A. V. Chubukov, Phys. Rev. B {\bf 43}, 3337 (1991).
\bibitem{akira} T. Hikihara, L. Kecke, T. Momoi, and A. Furusaki, \prb {\bf 78}, 144404 (2008).
\bibitem{sudan} J. Sudan, A. L\"uscher, and A. L\"auchli, \prb {\bf 80}, 140402(R) (2009).
\bibitem{mzh} M. E. Zhitomirsky and H. Tsunetsugu, Europhys. Lett. {\bf 92}, 37001 (2010).
\bibitem{syromyatnikov} A. V. Sizanov and A. V. Syromyatnikov, JETP Lett. {\bf 97}, 114 (2013).
\bibitem{khodas} I. I. Mazin, Phys. Rev. B 84, 024529 (2011), M. Khodas and A. V. Chubukov, Phys. Rev. Lett. 108, 247003 (2012).
\bibitem{sachdev2001} E. Demler, S. Sachdev, Y. Zhang, \prl {\bf 87}, 067202 (2001).
\bibitem{noz} P. Nozieres and D. Saint James, J. Physique, {\bf 43}, 1133 (1982); L. Radzihovsky, P. B. Weichman, J. I. Park,
Annals of Physics {\bf 323}, 2376 (2008).
\end{references}

\begin{references}
\bibitem{alicea_a} J. Alicea, A. V. Chubukov, and O. A. Starykh, Phys. Rev. Lett. {\bf 102}, 137201 (2009).
\bibitem{oguchi}
T. Oguchi, Phys. Rev. {\bf 117}, 117 (1960).
\bibitem{alicea_a2} C. Griset, S. Head, J. Alicea, O. A. Starykh, \prb {\bf 84}, 245108 (2011).
\end{references}
\end{document}